\documentclass[numbers,sort&compress]{IntechOpen-Book}
\usepackage{amsfonts}
\usepackage{amssymb}  
\usepackage{amsthm}
\usepackage{amsmath}
\usepackage{bbm}
\usepackage{multirow}
\usepackage{makecell}
\usepackage[bf,SL,BF]{subfigure}
\usepackage[english]{babel}
 \usepackage[justification=centering]{caption}

\usepackage{tikz}
\usepackage{verbatim}
\usepackage{pgf}
\usetikzlibrary{arrows,automata}
\usetikzlibrary{trees}
\usepackage[figuresright]{rotating}
\newcommand{\RNum}[1]{\uppercase\expandafter{\romannumeral #1\relax}}

\usepackage{lstbayes}
\lstdefinestyle{mystyle}{
    commentstyle=\color{DarkGreen},
    keywordstyle=\color{blue},
    stringstyle=\color{DarkGreen},
    basicstyle=\ttfamily\footnotesize,
    breakatwhitespace=false,         
    breaklines=true,                 
    frame=tb,
    keepspaces=true,                 
    numbers=left,                    
    numbersep=5pt,                  
    showspaces=false,                
    showstringspaces=false,
    showtabs=false,                  
    tabsize=2
}
\graphicspath{{Artworks/}}

\begin{document}

\Mainmatter

\begin{frontmatter}

\chapter{Dependent Dirichlet Processes for analysis of a Generalized Shared Frailty Model  }
\author{Chong Zhong}
\author{Zhihua Ma}
\author{Junshan Shen}
\author{Catherine Liu}

\makechaptertitle

\chaptermark{Dependent Dirichlet process}

\begin{abstract} 
Bayesian paradigm takes advantage of well fitting complicated survival models and feasible computing in survival analysis owing to the superiority in tackling the complex censoring scheme, compared with the frequentist paradigm.
In this chapter, we aim to display the latest tendency in Bayesian computing, in the sense of automating the posterior sampling, through Bayesian analysis of survival modeling for multivariate survival outcomes with complicated data structure. 
Motivated by relaxing the strong assumption of proportionality and the restriction of a common baseline population, we propose a generalized shared frailty model which includes both parametric and nonparametric frailty random effects
so as to incorporate both treatment-wise and temporal variation for multiple events. 
We develop a survival-function version of ANOVA dependent Dirichlet process to model the dependency among the baseline survival functions. 
The posterior sampling is implemented by No-U-Turn sampler in Stan, a contemporary Bayesian computing tool, automatically.
The proposed model is validated by analysis of the bladder cancer recurrences data. The estimation is consistent with existing results. 
Our model and Bayesian inference provide evidence that the Bayesian paradigm fosters complex modeling and feasible computing in survival analysis and Stan relaxes the posterior inference. 

\end{abstract}

\begin{keywords} 
ANOVA DDP, dependent treatments, multivariate survival outcomes, recurrence, Stan
\end{keywords}

\end{frontmatter}

\section{Introduction}
The shared frailty model, coined by \cite{vaupel1979impact}, has been widely used in the analysis of multivariate survival outcomes that might be associated within subgroups or clusters. 
Enormous work has been devoted to the development of shared frailty model in both Bayesian and frequency paradigms, and the reviews can be found in \cite{ibrahim2001bayesian, duchateau2007frailty, balan2020tutorial}.
As an extension of the well-known Cox's proportional hazard model, conditional on the frailty effect, the shared frailty model assumes the hazard ratio between two subjects is proportional to their difference in relative risk scores over time. In addition to the proportional hazard assumption, the shared frailty model fixes the baseline hazard function among all clusters. 

\begin{figure}[!htb]
\center
\subfigure[]{
\begin{minipage}[t]{0.45\linewidth}
\centering
\includegraphics[width=1\textwidth]{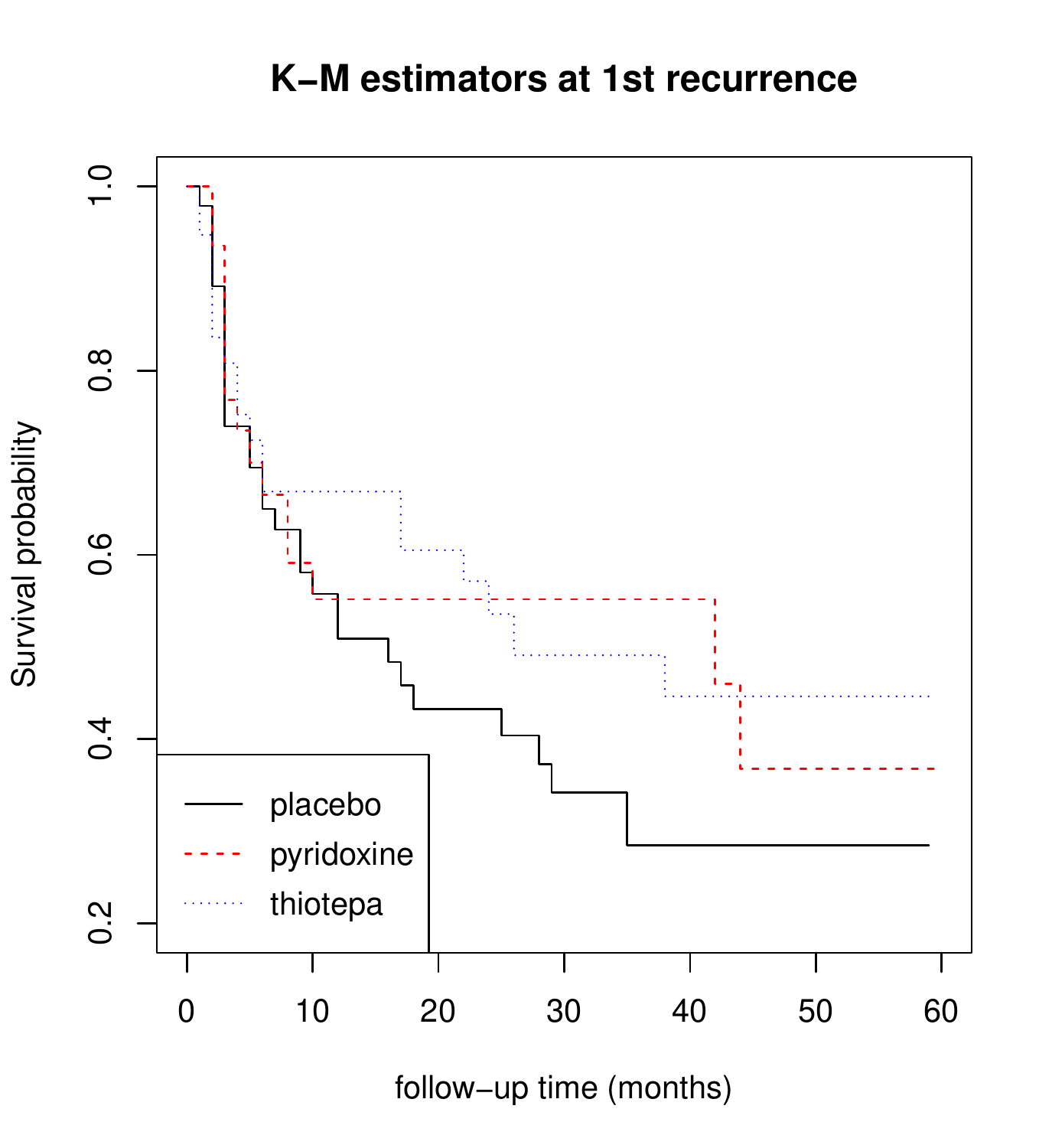}
\label{figKM11}
\end{minipage}%
}%
\subfigure[]{
\begin{minipage}[t]{0.45\linewidth}
\centering
\includegraphics[width=1\textwidth]{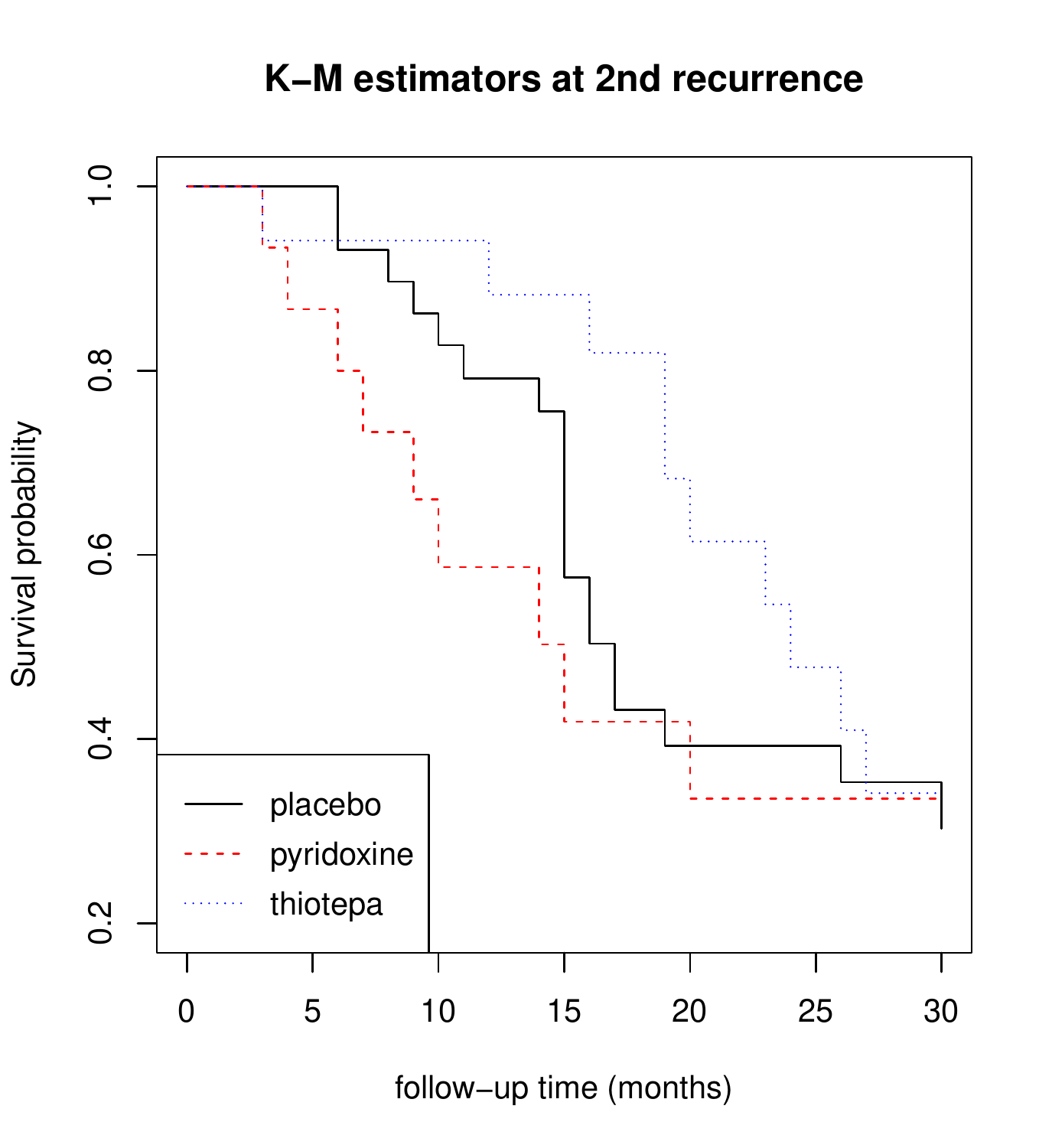}
\label{figKM2}
\end{minipage}%
}%
\caption{The Kaplan-Meier estimator of survival functions for first recurrence time (a) and second recurrence (b) in the bladder cancer data. }
\label{figBladder}
\end{figure}

Traditional shared frailty models provide a good framework for expediently mathematical tackling the heterogeneity among the multivariate observations, whereas in practice it needs modification and adaption to tolerate complex structure so as to incorporate cross information owing to the intra- and inter- subject variability (\cite{hanson2012bayesian, de2015bayesian}).
Take the renowned data on recurrences of bladder cancer for instance (\cite{survival-package}). There are three treatment arms, placebo, thiotepa, and pyridoxine. Patients had multiple recurrences of tumors which were sparse beyond the fourth recurrence. Figure \ref{figBladder} shows the Kaplan-Meier estimators of the survival function for the times of the first and the second recurrences under three treatments. One observes that, the estimated survival curves at the first recurrence are crossed indicating a crossed hazard and that the proportional hazard assumption is suspected (\cite{ZengLin2007JRSSBmaximum}); the survival curve of pyridoxine falls below that of placebo at the second recurrence compared to the first recurrence, indicating
the functional form of the survival curves varies between recurrences. 
Neglecting such characteristics of non-proportionality and stratification of recurrences may yield inefficiency by encumbering borrowing strength from potentially related information sources, and consequently may jeopardize the prediction of the global survival times.  
Moreover, dependency might be existing among the treatment strata and the stratification of recurrences (\cite{de2004anova,hanson2012bayesian}).

Consequently,  more complex modeling is needy to characterize the dependence among the baseline hazard functions and treatment strata due to the temporal effects of recurrences. 
Frequentist inference and computing are pretty challenging and even infeasible.
Existing Bayesian literature  considered modifications of shared frailty model based on some kind of partial aberrant phenomena  (\cite{de2014bayesian, paulon2020joint}; among others)
but rare work has taken bi-level stratification into account (\cite{conlon2014multi}), not to mention that dependence among treatment strata
(\cite{hanson2012bayesian}).

We propose a generalized shared frailty model (GSFM) for multiple events time data that allows the baseline hazard function to change along with the types of events and treatment strata, strengthening the ability to borrow information from many sources. 
The proposed model postulates multiplier frailty including both parametric and nonparametric ones, where the parametric frailty random effect accounts for the within subject association by treating each subject as a cluster; and a nonparametric frailty effect represents dependency among treatment strata and temporal recurrences. 
For the proposed model GSFM, we suggest a Bayesian solution to estimate the regression coefficient vector and the variance parameter of the frailty term, and baseline survival functions stratified by treatments and recurrences.
In a Bayesian workflow, the posterior distribution is determined by the combination of observational data in the form of likelihood function and the prior distribution represented based on the background knowledge. 
From a Bayesian perspective, we model the dependent nonparametric prior through transferring the data context aforementioned into the ANOVA dependent Dirichlet process (ANOVA DDP), which will be further reviewed in Section 2. 
The construction of No-U-Turn sampler for Markov chain Monte Carlo (MCMC) sampling is automated by Stan (\cite{standev2021stancore}) with its R interface (\cite{Rstan}). The posterior inference is conducted by Stan as well. 

The rest of this chapter is organized as follows. In Section 2, under typical data scenarios of dependence structure, we summarize several modification versions of the dependent Dirichlet process (DDP) initiated from MacEachern's regression spirit that nested dependent predictors into the traditional Dirichlet Process (DP). In Section 3, we postulate the GSFM and transform the dependent dual-stratified multiple events to the survival-function based version of the ANOVA DDP. We have a short comparison between Stan and Nimble, two contemporary Bayesian computing tools based on our user experience. In section 4, we demonstrate the validity of the GSFM and Bayesian inference and analysis of the data on recurrences of bladder cancer. A brief conclusion is contained in Section 5. 

\section{Review of MacEachern's DDP}
The DP is the most popular Bayesian nonparametric prior since the seminal work of \cite{ferguson1974prior}. 
The belief in data background that there exists some kind of dependence structure stimulates construction and selection of proper dependent prior.
Some dependent DPs are constructed for unsupervised purposes such as clustering (\cite{teh2006hierarchical, rodriguez2008nested}). The DDP prior adopted in our proposed model 
is supervised and predictor-dependent, originated from \cite{maceachern2016nonparametric, quintana2020dependent}, named as MacEachern's DDP in two recent review papers, which are interpretive and comprehensive (\cite{maceachern2016nonparametric, quintana2020dependent}). The key idea behind the MacEachern's DDP is that the distributions of the random measures are marginally DP distributed, validated by in our subsections 3.2 and 3.3. 
Therefore we here confine how the MacEachern's DDP (henceforth we use the DDP to denote the MacEachern's DDP if the context is clear) came into being expanded from the DP, and compare various modification versions of the DDP under various dependent data structures.
\\
~\\ 
{\textit{DP vs. DDP}}\\
The DP is a distribution on distributions whereas the DDP aims to construct prior for a collection of distributions $\mathcal{F} = \{F_x|x\in \mathcal{X}\}$ indexed by covariate $x$. 
In general, there are several representations of the DP such as Polya Urn, Levy measure, and stick-breaking representations (\cite{phadia2015prior}). 
Here we use Sethuraman’s stick-breaking construction to connect the DP with the DDP. The stick-breaking construction is a kind of infinite sum representation that divides the DP into two countable series, the weights and the atoms. Generally, a DP is expressed as a process with two components, the mass parameter determining the weights and the base measure to generate atoms. 
Through the stick-breaking construction, the DDP can be easily extended from the DP. We list their comparison in Table \ref{Tab:Comparsion}, where we can find that the dependency among the covariates set $\mathcal{X}$ is realized by indexing the mass parameter and base measure with the covariate $x\in \mathcal{X}$. More specifically, the dependency can be characterized through the dependency among the weights and atoms in the DDP. 
\begin{table}[!htb]
\caption{Comparison of DP \& DDP}
\begin{tabular}{lp{5cm}p{5cm}}
  & {\color{blue}DP}   &{\color{blue} DDP}  \\  \vspace{0.5cm}
{\color{blue} RPM}   & $F \sim \text{DP}(M, F_0)$ & $\mathcal{F} =\{F_x|x\in \mathcal{X}, M_x, F_{0x}\}$  \\ \vspace{0.5cm}
\makecell[l]{\color{blue} Sethuraman’s \\ \color{blue} construction }& \makecell[l]
 {$F(\cdot) = \sum_{h=1}^{\infty}p_h \delta_{\theta_h}(\cdot)$ \\ $p_h \sim \text{SBW}(1, M)$\\ $\theta_h \sim F_0$}  &  \makecell[l]{$F_x(\cdot) = \sum_{h=1}^{\infty}p_{xh}\delta_{\theta_{xh}}(\cdot)$\\ $p_{xh} \sim \text{SBW}(1, M_x)$ \\ $\theta_{xh} \sim F_{0x}$}\\ \vspace{0.5cm}
{\color{blue} Convolution}   & $H(y) = \int k(y|\theta)dF(\theta)$ & $H_x(y) = \int k(y|\theta)dF_x(\theta)$  \\ \vspace{0.5cm}
\end{tabular}
\label{Tab:Comparsion}
\end{table} 

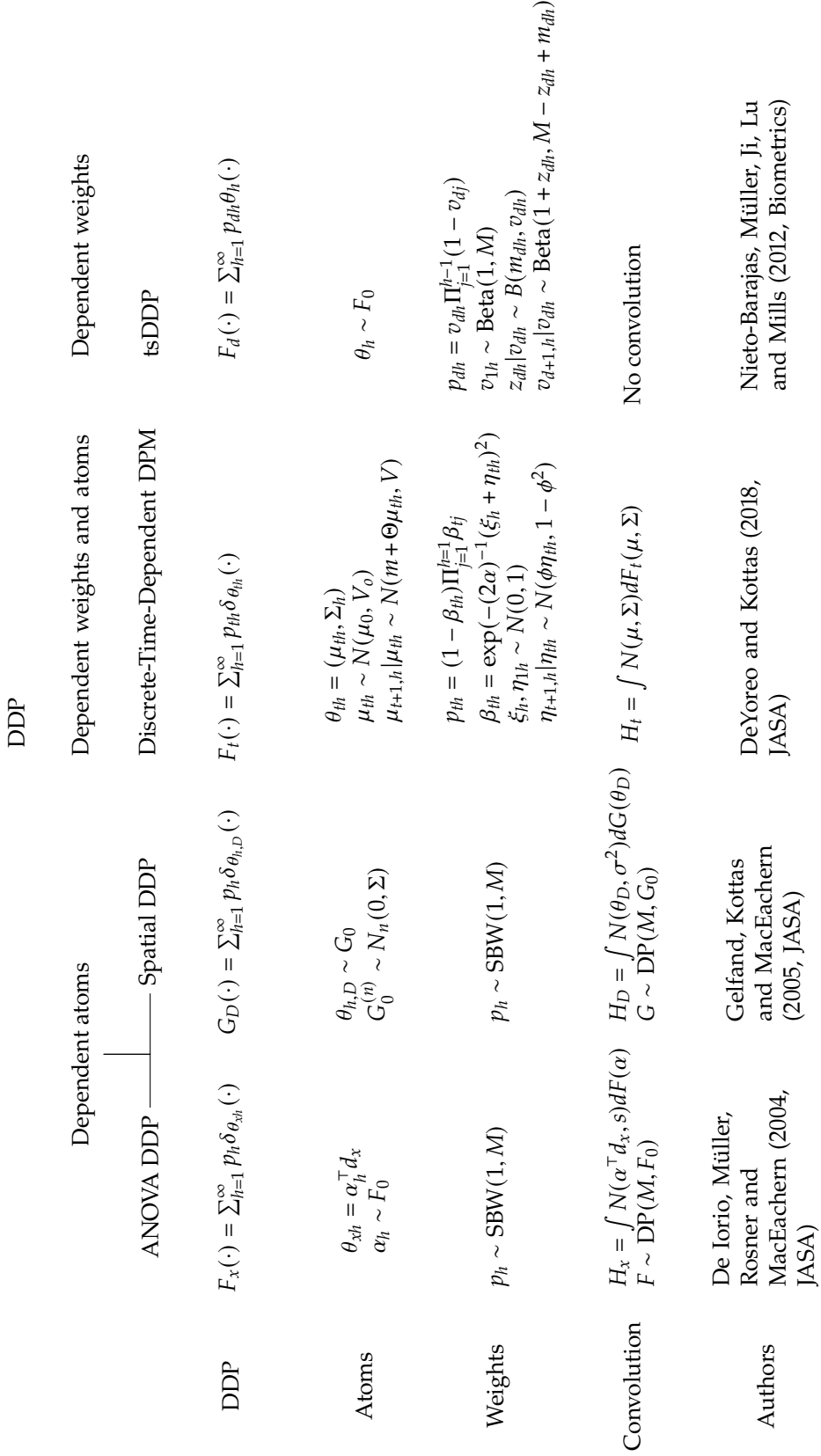
\begin{sidewaysfigure}[p]
\begin{tikzpicture}
\node (DDP) {DDP};
\node[ below of=DDP,yshift=-1cm] (Atoms) {Atoms};
\node[ below of=Atoms,yshift=-1cm] (Weights) {Weights};
\node[ below of=Weights,yshift=-1cm] (convolution) {Convolution};

\node[  right of=DDP, xshift=2cm] (DDP1) {$F_x(\cdot) = \sum_{h=1}^{\infty}p_{h}\delta_{\theta_{xh}}(\cdot)$};
\node[  right of=DDP1, xshift=3cm] (DDP2) {$G_D(\cdot) = \sum_{h=1}^{\infty}p_{h}\delta_{\theta_{h, D}}(\cdot)$};
\node[  above of=DDP1, yshift=0.2cm] (ANOVA DDP) {ANOVA DDP};
\node[  above of=DDP2, yshift=0.2cm] (Spatial DDP) {Spatial DDP};
\node[  above of=ANOVA DDP, xshift=2cm] (DA) {Dependent atoms};

\draw (DA) |- node[anchor=north] { } (ANOVA DDP);
\draw (DA) |- node[anchor=north] { } (Spatial DDP);

\node[  right of=Atoms, xshift=2cm, text width=2cm,align=left] (Atoms1) {$\theta_{xh}=\alpha_h^{\top}{d_x}$ \\ $\alpha_h \sim F_0$};
\node[  right of=Atoms1, xshift=3cm, text width=3cm,align=left]  (Atoms2) {$\theta_{h, D} \sim G_0$ \\ $G_0^{(n)} \sim N_n(0,\Sigma)$};

\node[  right of=Weights, xshift=2cm, text width=3cm,align=left] (Weights1) {$p_h \sim \text{SBW}(1,M)$};
\node[  right of=Weights1, xshift=3cm, text width=3cm,align=left]  (Weights2) {$p_h \sim \text{SBW}(1,M)$};

\node[  right of=convolution, xshift=2.5cm, text width=4cm,align=left] (convolution1) {$H_x=\int N(\alpha^{\top} d_x, s)dF(\alpha)$\\ $F\sim$ DP$(M,F_0)$};
\node[  right of=convolution1, xshift=3cm, text width=4cm,align=left]  (convolution2) {$H_D=\int N(\theta_D,\sigma^2)dG(\theta_D)$\\ $G\sim$ DP$(M,G_0)$};

\node[  right of=DA, xshift=6cm,text width=5cm] (DWA) {Dependent weights and atoms};
\node[  right of=Spatial DDP, xshift=4cm,text width=5cm] (DtdDPM) {Discrete-Time-Dependent DPM};
\node[  right of=DDP2,xshift=3cm] (DDP3) {$F_t(\cdot)=\sum_{h=1}^{\infty}p_{th}\delta_{\theta_{th}}(\cdot) $};
\node[  right of=Atoms2,xshift=4cm,text width=4cm] (Atoms3) {$\theta_{th}=(\mu_{th},\Sigma_h)$\\
$\mu_{th} \sim N(\mu_0,V_o)$\\
$\mu_{t+1,h}| \mu_{th} \sim N(m+\Theta \mu_{th},V)$
};
\node[  right of=Weights2,xshift=5cm,text width=6cm] (Weights3) {$p_{th}=(1-\beta_{th})\Pi_{j=1}^{h=1}\beta_{tj}$\\
$\beta_{th}=\text{exp}(-(2\alpha)^{-1}(\xi_h+\eta_{th})^2)$\\
$\xi_{h},\eta_{1h}\sim N(0,1)$\\
$\eta_{t+1,h}|\eta_{th} \sim N(\phi \eta_{th},1-\phi^2)$
};

\node[  right of=convolution2,xshift=3cm] (convolution3) {$H_t=\int N(\mu,\Sigma) d F_t(\mu,\Sigma)$};

\node[  right of=DA, xshift=11.5cm,text width=4cm] (DW) {Dependent weights};
\node[  right of=DtdDPM,xshift=4cm, text width=3cm] (ts DDP) {tsDDP};
\node[  right of=DDP3,xshift=5cm] (DDP4) {$F_d(\cdot) = \sum_{h=1}^{
\infty}p_{dh}\theta_h(\cdot)$};
\node[ right of=Atoms3,xshift=3cm] (Atoms4) {$\theta_{h}\sim F_0$};
\node[  right of=Weights3,xshift=4cm,text width=6cm] (Weights4) {$p_{dh}=v_{dh}\Pi _{j=1}^{h-1}(1-v_{dj})$\\
$v_{1h}\sim \text{Beta}(1,M)$\\
$z_{dh}| v_{dh}\sim B(m_{dh},v_{dh})$\\
$v_{d+1,h} |v_{dh}\sim \text{Beta} (1+z_{dh},M-z_{dh}+m_{dh})$

};

\node[  right of=convolution3,xshift=3.5cm] (convolution4) {No convolution};

\node[  below of=convolution, yshift=-1cm] (Authors) {Authors};
\node[  right of=Authors,  xshift=2cm,text width=3cm] (Authors1) {De Iorio, Müller, Rosner and MacEachern (2004, JASA)};
\node[  right of=Authors1, xshift=3cm,text width=3cm] (Authors2) {Gelfand, Kottas and MacEachern (2005, JASA)};
\node[  right of=Authors2,xshift=4cm, text width=5cm] (Authors3) {DeYoreo and Kottas (2018, JASA)};
\node[  right of=Authors3, xshift=4.5cm,text width=5cm] (Authors4) {Nieto‐Barajas, Müller, Ji, Lu and Mills (2012, Biometrics)};

\node[  above of=DA,xshift=5cm] (DDP) {DDP};

\end{tikzpicture}
\caption{Workflows of representative expansions of DDP} 
\label{fig:DDPmodels}
\end{sidewaysfigure}

The DDP can be widely applied to scenarios of various dependence data structure. 
We review modification versions of the DDP from three categories depending on which part it modifies in the stick-breaking representation, weights, atoms, or both. 
 The first is to impose the dependency on the atoms but keep common weights, leading to two typical representatives, ANOVA and Spatial (\cite{de2004anova, gelfand2005bayesian, de2009bayesian}).  The ANOVA type DDP encoded the covariate dependence in the form of regression  for the atom processes.  The Spatial DDP models for nonstrationary spatial random fields with heterogeneous variance.
The second category is to modify the weights to be dependent but keep the common atoms. The early and typical work is the time series DDP (\cite{nieto2012time}). They introduced a Markov Beta process on the weights to account for the temporal dependency. 
The third category is to impose dependency on both weights and atoms (\cite{deyoreo2018modeling}). They constructed vector autoregressive and autoregressive models for atoms and weights, respectively. We summarize the aforementioned types of typical modifications in Figure \ref{fig:DDPmodels}.

\section{Model and Bayesian inference}

Consider a clinical trial with multiple event types, for example, the time of the $k$th recurrence of a certain disease. In the trial, $n$ subjects are divided into $G$ strata of treatment. Our goal is to describe the relationship between the time to the $k$th recurrence of a subject, and its treatment stratum as well as its vector covariates Z.
For a certain subject, the times of recurrences may be dependent since they occur on the same individual and thus we assume an unobservable independent shared-frailty random effect $W$ to account for this dependence. 
On the other hand, we may allow the conditional hazard affiliated with the script pair $kj$ implying distinct survival distributions along with the temporal order of the recurrences of the disease and for specific treatment.
For the $i$th subject in the $j$th treatment stratum, at the $k$th recurrence, given the value of frailty variable $w_i$ and its covariate vector $z_{kji}$, we propose the following frailty model,
\begin{align}
    \label{basicmod}
    \lambda_{kj}(t|w_i, z_{kji}) = w_i \lambda_{0kj}(t)\exp(\beta^T z_{kji}), k=1, \cdots, K, j=1, \cdots, G, i = 1, \cdots, n_j.
\end{align}
Model \eqref{basicmod} is called the \emph{generalized} shared frailty model in the sense that non-proportionality among $k$-varying recurrences is allowed by the fact that the right-hand baseline hazard has footnotes $k$ and $j$. We allow dependency among treatment strata in model \eqref{basicmod}.
Therefore, the baseline hazard function $\lambda_{0kj}$ acts as if a nonparametric frailty random measure accounting for the dependency owing to the recurrences and treatment scheme.

 Model \eqref{basicmod} is an extension of the classical shared frailty model (4.1.1) on page 101 of \cite{ibrahim2001bayesian} since the baseline hazard function there does not vary from the recurrences and the treatment strata. 
Model \eqref{basicmod} has the analog spirit to the frailty model (1) in \cite{de2014bayesian}, whereas their treatment strata are independent.

\subsection{Likelihood}

The corresponding survival function of model \eqref{basicmod}  is given by:
\begin{align*}
    S_{kj}(t|w_i, z_{kji}) = \{S_{0kj}(t)\}^{\exp(\beta^T z_{kji} + v_i)},
\end{align*}
where $S_{0kj}$ denotes the baseline survival function of the $k$th recurrence for subjects in the $j$th treatment stratum, $v_i = \log(w_i)$ denotes logarithm transformation of the frailty effect. Let $f_{0kj}$ be the corresponding baseline density function. 

Given the data sample  $(Y_{kji}, \delta_{kji}, z_{ji})$,
where $Y_{kji} = \min(C_{kji}, T_{kji})$, $\delta_{kji} = I(T_{kji} \le C_{kji})$, with 
$T_{kji}$ being the gap time between the $(k-1)$th and $k$th recurrence of the $i$th subject in the $j$th stratum and $C_{kij}$ being the corresponding censoring variable that is independent of $T_{kji}$ given the covariate vector $z_{kji}$, for $k=1, \cdots, K, j=1, \cdots, G, i = 1, \cdots, n_j$, and $\sum_{j=1}^{G}n_j = n$. 
  In the $j$th stratum, suppose that there are $n_{kj}$ ($n_{kj} \le n_j$)  subjects suffering from the $k$th recurrence. Then the likelihood is written as:
\begin{align*}
   \prod_{k=1}^{K}\prod_{j=1}^{G}\prod_{i=1}^{n_{kj}} 
    &[\exp(\beta^T z_{kji} + v_i) f_{0kj}(y_{kji}) \{S_{0kj}(y_{kji})\}^{(\exp(\beta^T z_{kji} + v_i)-1)}]^{\delta_{kji}}\\
    & \times \{S_{0kj}(y_{kji})\}^{(1-\delta_{kji})\exp(\beta^T z_{kji} + v_i)}.
\end{align*}

\subsection{Survival-function based version of the ANOVA DDP}
In the Bayesian workflow for the estimation, prior distributions are first determined. We here specify appropriate nonparametric priors for $S_{0kj}$ and $f_{0kj}$. Since they can be easily derived from one to the other, we here only introduce the priors for $S_{0kj}$. 

We divide $S_{0kj}$ into $K$ groups, and the $k$th group has $G$ baseline survival functions of different treatment strata at the $k$th time of recurrence.  That is, for a fixed $k$, $\mathcal{S}_k = \{S_{0kj}, j=1, \cdots, G\}$ is a collection of baseline survival functions with length $G$ indexed by the categorical covariate $j$ denoting the treatment stratum. The next procedures come from the spirit of \cite{de2004anova}. 
As a general example, suppose two dugs $A$ and $B$ will be taken in treatment, with $V$ and $U$ levels of doses, respectively. In this case, $G=VU$ denotes the number of treatment strata and let the level of the $j$th stratum be $(v, w)$. We write the stick-breaking form of $S_{0kj}$ such that $S_{0kj}(t) = \sum_{h=1}^{\infty}p_h I(t>\theta_{kjh})$. We impose an ANOVA structure on $\theta_{kjh}$ :
\begin{align}
\label{anova}
    \theta_{kjh} = m_{kh} + A_{kvh} + B_{kwh},
\end{align}
where $m_{kh}$ denotes the ANOVA effect shared by all the strata at the $k$th recurrence, and the rest terms are the ANOVA effects of the $j$th stratum at the $k$th recurrence. 
 Let the three components be independently generated from three distributions, and marginally on $j$, {the baseline survival function $S_{0kj}$ follows a DP}. The aforementioned procedure implies that $\mathcal{S}_k$ is a survival-function version of the ANOVA DDP. 

Since any function in the stick-breaking form is discrete almost surely, we place a convolution through the Dirichlet process mixture (DPM) model (\cite{lo1984class}). Particularly, since the baseline survival functions are defined on the positive half real line, the convolution kernel in DPM should be positive such as log-normal, Gamma, and Weibull. In this chapter, a log-normal kernel is considered. For different recurrences, we treat the relationship among $\mathcal{S}_k$s to be independent. 

\subsection{One-way ANOVA DDP}
Considering the data of our interest, where only one drug and one level of dose is used in each treatment stratum, we introduce the modeling of the survival-function version of one-way ANOVA DDP.  
In this case the prior for the $\mathcal{S}_k$ reduces to a one-way ANOVA form since the dependency among the $G$ treatment strata is explained by only one ANOVA effect. Furthermore, if we set $m_{kh}=0$ , $\alpha_{kh} = (\theta_{k1h}, \cdots, \theta_{kGh})^T$ reduces to a $G$-variate variable denoting the locations of all $G$ baseline distributions and thus $\theta_{kjh} = \alpha_{kh}^T d_j$, where $d_j$ is the design vector of the $j$th stratum to select the appropriate ANOVA effects corresponding to $j$. 

With the above notations, we summarize the procedure to construct the survival-function version of one-way ANOVA DDP prior in model \eqref{basicmod} as follows: 
\begin{enumerate}
    \item Stick-breaking form. For $k=1, \cdots, K$,  let $\mathcal{H}_k$ be the collection of $G$ distribution functions s.t $\mathcal{H}_k=\{H_{kj}, j=1, \cdots, G\} $.
    $H_{kj}(\cdot) = \sum_{h=1}^{\infty} p_{kh} \delta_{\theta_{kjh}}(\cdot)$.
 
    \item Convolution step.  Let $\alpha_{kh} = (\theta_{k1h}, \cdots, \theta_{kGh})^T$, and $d_j$ be the $j$th design vector of length $G$ with the $j$th element being $1$ and others being $0$. 
    Let $H_{0k} = (H_{0k1}, \cdots, H_{0kG})$ be the collection of base measures, 
    $S_{0kj}(t) = \int S_{\text{LN}}(t|\alpha_k^T d_j, \sigma^2) dH_k(\alpha, \sigma)$, where $S_{\text{LN}}$ denotes the survival function of the log-normal distribution, and $H_k \sim \text{DP}(M_k, H_{0k})$. 
    
    \item Determine the mass parameter and the base measure. For simplicity, we set $M_k = 1$ for all $k$, which is a commonly used default value of the mass parameter (\cite{gelman2013bayesian}) , $H_{0k}(\theta, \sigma) = N(0, I_{G}) \times \text{Cauchy}(0, 5)^+ $, where $\text{Cauchy}^+$ denotes the half\_Cauchy distribution. 
    
\end{enumerate}
Step 1 is a standard stick-breaking representation for DP. Step 2 is kernel mixture of DP whereas the kernel is a survival function rather than a cumulative distribution function. The realization of Step 2 is quite straightforward in Stan as it provides the function \texttt{lognormal\_lccdf} to be used as the kernel of the survival function of the log-normal family. 

In Step 3, we specify the base measure as the prior for the location and shape parameters of the log-normal kernel directly rather than adding another hyper prior distribution like \cite{de2009bayesian} did. The main reason is to simplify the computation in Stan. Particularly, inspired by \cite{gelman2006prior} and \cite{gelman2008weakly}, we use the half-Cauchy distribution as the non-informative prior for the variance parameter instead of the inverse Gamma prior. In our practice, the choice of half-Cauchy prior significantly improves the speed of convergence and mixture performance of the MCMC chains in our real data analysis and simulation. Another interesting point we met in numerical studies is that the informativeness of the base measure for $\theta$. Here we don't assign the non-informative distribution but a weakly informative one is considered since we find such a weakly informative prior provides better MCMC performance than that of non-informative one with higher effective sample size and better mixture performance. In our other research experience, the weakly informative prior for the variance parameter in the mixing component of the DPM seems to be more preferable. 

\subsection{Other priors and MCMC}

In terms of the prior for the parametric prior $w_i$, we choose log normal prior that $v_i = \log(w_i)$ and $v_i \sim N(0, \tau^2)$, where $\tau>0$ is an unknown parameter. We further assign a half Cauchy prior for $\tau$ s.t $\tau \sim \text{Cauchy}^+(0, 5)$ as a non-informative prior. 
The prior for the vector of regression coefficients is $\beta \sim N(0, 1000I)$ as a non-informative prior. 

We use the truncated Dirichlet process to replace the infinite summand in the DP. The selection of the truncation point is often ad-hoc. Since in Stan the NUTS cannot sampler discrete parameters, we have to fit the truncation number and the mass parameter before the MCMC procedure. In general, the truncation number is set to be large enough s.t the truncated part is negligible. \cite{ohlssen2007flexible} suggests to use a truncation number $L$ that is greater than $5M+5$. In our computation, we set $L=12$. 

The MCMC sampling for the posterior distribution is realized in Stan. Stan and its \textbf{R} version are widely used in statistical modeling and high-performance statistical computing, especially in Bayesian. Stan realizes the MCMC sampling through the No-U-Turn sampler (NUTS). Stan automates the deriving of the full conditional posterior distribution and NUTS is able to obtain high effective sample size (\citep{hoffman2014no}). 

\subsection{Stan and NIMBLE: programming styles}
The MCMC sampling procedure is implemented in Stan and we also tried to implement the model in NIMBLE, another contemporary Bayesian computing tool in \textbf{R}.
Stan and NIMBLE are two contemporary Bayesian computing tools that have drawn arising interest for Bayesian analysis but still remain under active development (\cite{kerioui2020bayesian, ma2021bayesian}). 
The main advantage of Stan and NIMBLE is that they provide clear automatic posterior sampling procedures based on their specific sampling algorithms without particular justification. Therefore, users can be released from complicated probabilistic deriving and implementation.
There has been buzz group discussion about the comparison between Stan and NIMBLE in environments like \cite{Stanforum} and \cite{NIMBLEusers}. One comparison on their built-in samplers is demonstrated through implementing weakly informative and informative estimation within the trimmed mean regression model setting (\cite{Zhang2021ABayesian}). 
Here we contribute a naive comparison on their programming styles based on the first two authors' experience in coding this project and using Stan and NIMBLE, respectively. 

A Bayesian paradigm is made up of three main steps, the prior, likelihood, and the posterior. MCMC generates samples to approximate the posterior distribution. Therefore, what one needs to set in a Bayesian computing tool is the prior and likelihood, let alone Stan or NIMBLE. Nevertheless, Stan and NIMBLE take different programming styles in writing likelihood. In Stan, the default way to present the log likelihood is the syntax \texttt{target} and users can add log contribution to it freely, which is similar to the natural language and straightforward to users whatever level of mathematical background. In NIMBLE, the default way is to transfer the likelihood into some standard distributions given by NIMBLE, which may not be friendly for users who have a relatively less mathematical background. 

We take fitting the finite mixture of Gaussian model as an example. For a fixed positive integer $L$, the distribution of $Y$ is given by $F_{Y}(s) = \sum_{l=1}^{L}p_l N(s|\mu_l, \sigma_l^2)$ and the log-likelihood is 
$\log L(p, \mu, \sigma |Y) = \sum_{i=1}^{n}\sum_{l=1}^{L}\{\log(p_l)+ \log \phi(y_i|\mu_1, \sigma_l)\}$, where $\phi$ denotes the density function of normal distribution. 
 The code for Stan and NIMBLE to implement this model is listed in Listing 1.1 and 1.2, respectively. In Listing 1.1 we clearly find that the contribution to the syntax \texttt{target} is just the sum of $\log(p_l)$ and the logarithm of the density of normal distribution denoted by \texttt{normal\_lpdf}. The rest is to assign a Dirichlet prior for the weights $p_l$ and other parameters. However, in NIMBLE code shown in Listing 1.2, we have to transfer the likelihood into some sampling procedures by IMAGING that there are $L$ clusters of random numbers, the random numbers are i.i.d Gaussian within each cluster, and the probability a random number is drawn from the $l$th cluster is $p_l$. Thereafter, the Dirichlet prior is assigned to $p_l$s. 
 Such imagine matches the Bayesian philosophy but when the likelihood function becomes to be quite complicated, to understand this sampling procedure may not be easy anymore, especially for practitioners not coming from a mathematics or statistics background.

\begin{lstlisting}[language=Stan,label=Stan,caption=Stan code for modeling mixture of Gaussian distribution]
data{
int<lower=1> N;
vector[N] y;
int<lower=1> L;
}
parameters{
simplex[L] p;
vector[L] mu;
vector<lower=0>[L] sigma;
}
model{
p ~ dirichlet(rep_vector(1, L));
mu ~ normal(0, 100);
sigma ~ cauchy(0, 2.5);
for(i in 1:N){
vector[L] lp_i;
for(l in 1:L){
lp_i[l] = log(p[l]) + normal_lpdf(y[i]|mu[l], sigma[l]);
}
target += log_sum_exp(lp_i);
}
}

\end{lstlisting}
~\\

\begin{lstlisting}[language=R,label=NIMBLE,caption=NIMBLE code for modeling mixture of Gaussian distribution]
NimbleCode <- nimbleCode({
  for (i in 1:N) {
    y[i] ~ dnorm(mu_y[z[i]], tau = tau_y[z[i]])
    z[i] ~ dcat(p[1:L])
  }
  for (j in 1:L) {
    mu_y[j] ~ dnorm(0, 0.01)
    tau_y[j] ~ dgamma(0.01, 0.01)
  }
  p[1:L] ~ ddirch(alpha0[1:L])
})
NimbleData <- list(y = y)
NimbleConsts <- list(L = L, N = length(NimbleData$y), alpha0 = rep(1, L))
NimbleInits <- list(mu_y = rnorm(NimbleConsts$L), tau_y = rgamma(NimbleConsts$L),p = rep(1/NimbleConsts$L, NimbleConsts$L))

\end{lstlisting}
 

\section{Application: bladder cancer recurrences}
We apply the GSFM to analyze the Bladder cancer recurrences data set contained in R package \texttt{survival}. Totally 118 subjects in the clinical trial are divided into $3$ treatment strata including placebo, pyridoxine (vitamin B6), and thiotepa. Each subject may experience $k$ (from 1 to 9) times of recurrences and may die from or not from the recurrence of bladder cancer. We don't discriminate the death from cancer and the recurrence, and the death from other causes is treated as censoring status. Our interest is the gap time between the $(k-1)$th and the $k$th recurrences. 
Besides the treatment schemes, two clinical covariates are considered: the number of tumors at the beginning $(x_1)$ and the size of the largest tumor $(x_2)$ within a subject. The values of these two covariates are evaluated at the beginning of each recurrence interval. 
This data set was once analyzed for the time between the first to the second recurrence as a univariate time-to-event outcome (\cite{zeng2006efficient}). 
In this chapter, we consider both the first and the second recurrences and thus $K=2$ here. The two covariates are scaled by divided by $100$. To simplify the computation, the follow-up time is transferred from months to years to get lower scalars. 

\subsection{Model checking for baseline survival functions}
\begin{figure}[!htb]
\centering
\subfigure[]{
\begin{minipage}[t]{0.45\linewidth}
\centering
\includegraphics[width=1\textwidth]{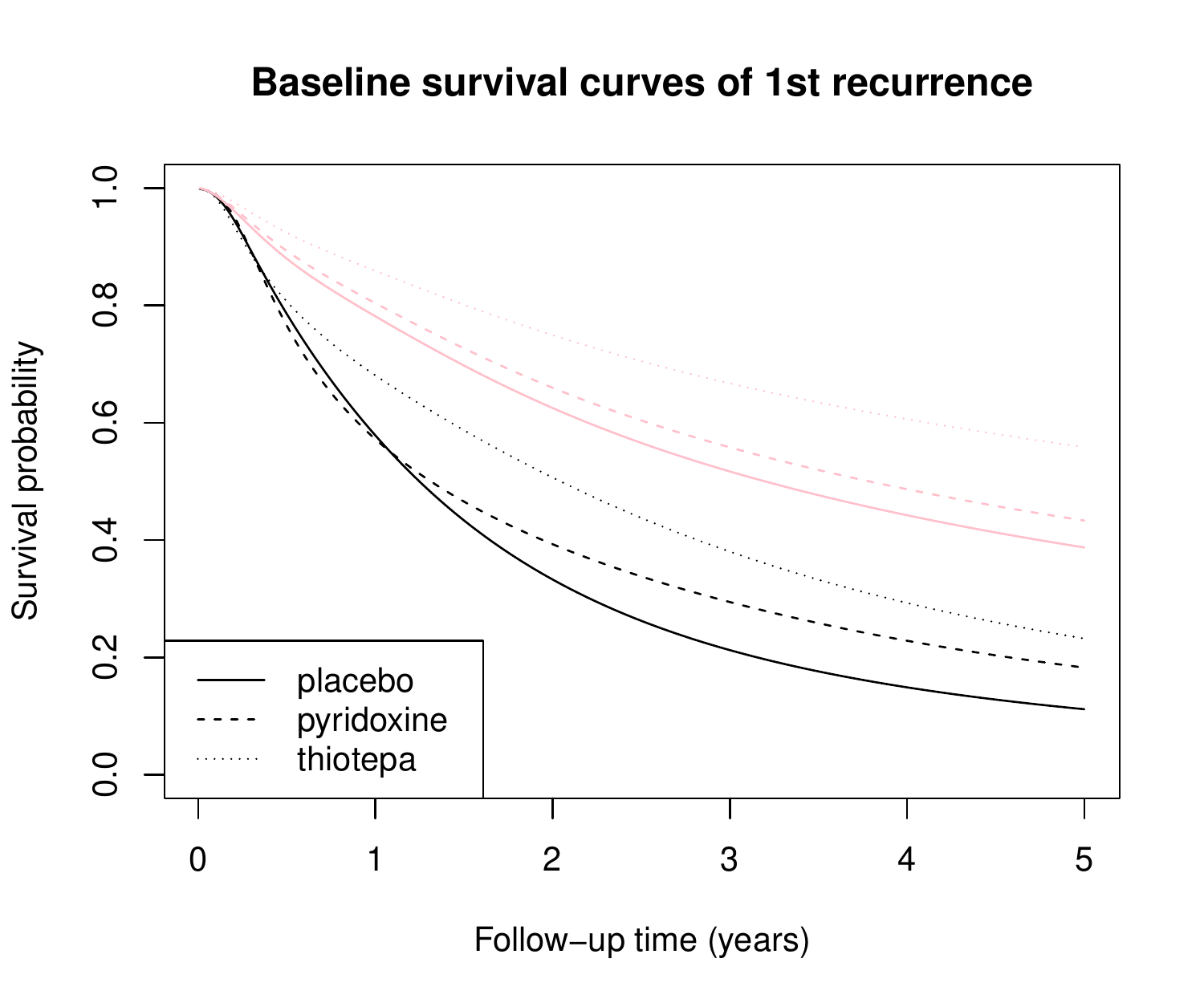}
\label{figRec1}
\end{minipage}%
}%
\subfigure[]{
\begin{minipage}[t]{0.45\linewidth}
\centering
\includegraphics[width=1\textwidth]{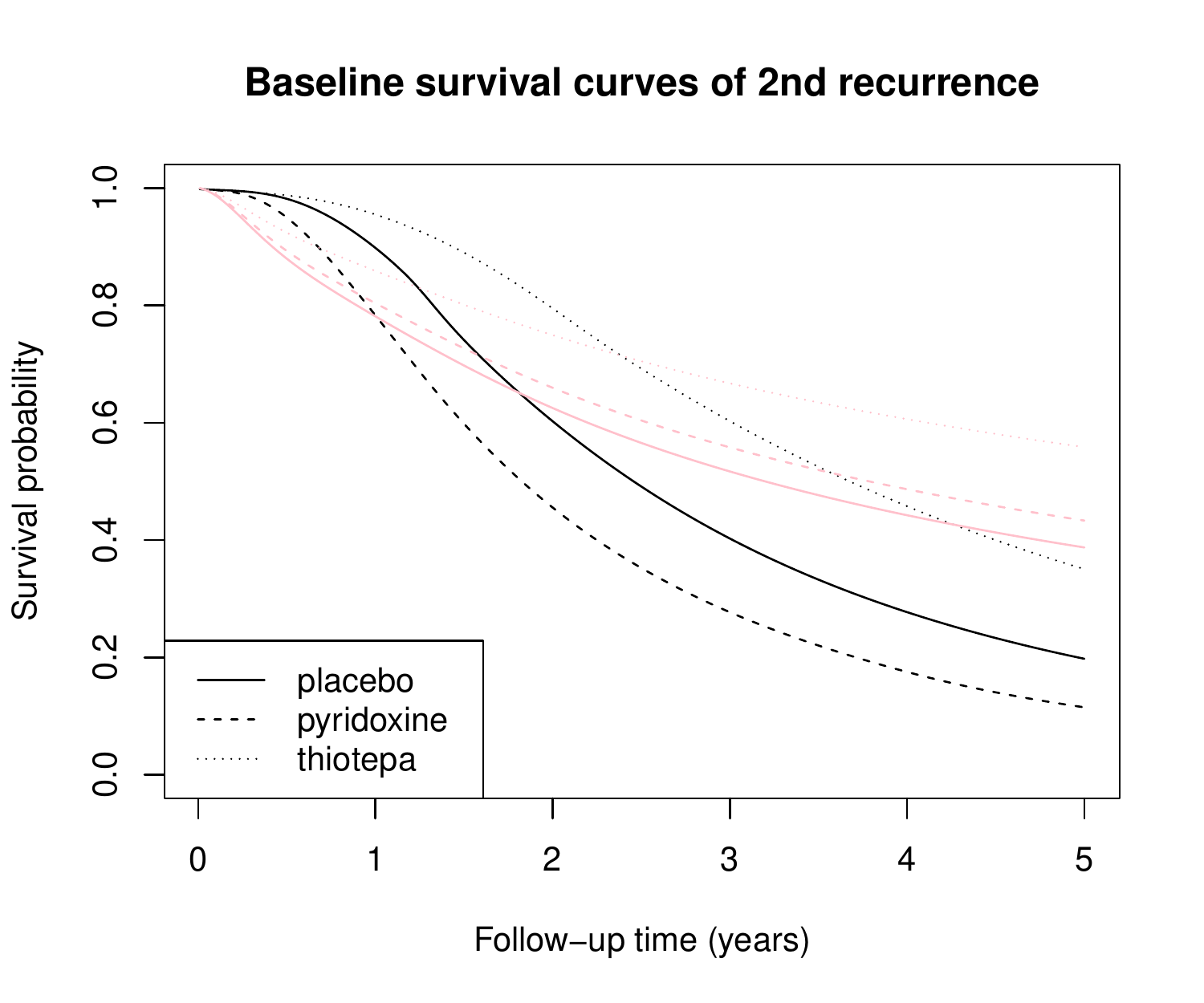}
\label{figRec2}
\end{minipage}%
}%
\caption{\footnotesize{The estimated baseline survival curves for the first (a) and second (b) recurrence; the black curves are estimated under the proposed generalized shared frailty model, and the pink curves are estimated under the traditional shared frailty model; the real lines, placebo; the dash lines, pyridoxine; the dotted lines, thiotepa. }}
 \label{figApp}
\end{figure}
Before further inference, we need to check whether the proposed model is appropriate. As an alternative, a shared frailty model is fit by R package \texttt{spBayeSurv}. In the shared frailty model, the treatment strata are considered as indicator covariates in the parametric term. We run $4$ independent MCMC chains for 5000 times with the first 2000 times burn-in and aggregate the rest chains together as the posterior samples under the GSFM. All chains are well mixed and convergent under the GSFM. 
For the shared frailty model, we run the MCMC 16000 times with the first 6000 times burn-in through R function \texttt{survregbayes} using the ``IID" Gaussian frailty under ``PH" model name. Other settings are default. 

The plots of the estimated baseline survival functions under different models stratified by treatment strata can be viewed in figure \ref{figApp}. From that, we find the baseline survival functions estimated under the GSFM shows similar trends as that of the K-M estimator in each recurrence, and reflects the
crossing survival curves at the first recurrence like the K-M estimator. 
However, the curves estimated by the shared frailty model are not crossed and cannot change along with recurrences.
Therefore, the proposed GSFM is appropriate for the data.

\subsection{Parametric estimation \RNum{1}: real data}
We use the mean of posterior samples (median for $\tau$) as the estimator of parameter and we list the estimation of vector of regression coefficients $\beta$ and standard deviation parameter $\tau$ in Table \ref{tab:realdata}. 

\begin{table}[!htb]
    \centering
        \caption{The parametric estimation and the MCMC performance for the bladder cancer recurrences data. Est, point estimation; SD, posterior standard deviation; ESS, effective sample size; PACE, the MCMC Pace. }

    \begin{tabular}{ccccc}
      & Est  & SD & ESS & PACE\\
 No. tumours &  13.849 &  11.051 & 1495 & 0.145\\
Tumour size & -14.196 & 12.341 & 1114 & 0.194 \\
 $\tau$ & 1.793 & 0.383 & 456 & 0.474
    \end{tabular}
    \label{tab:realdata}
\end{table}
From table \ref{tab:realdata} we find that as the number of tumors at start point increases, the hazard for recurrences increases as well whereas the larger size of the largest tumor will decrease the hazard. This conclusion is similar to that in \cite{zeng2006efficient} who analyzed the first recurrence as a univariate time-to-event data by a transformation model. 

Besides the parametric estimation result, we also report two metrics about the MCMC performance here. The first one is the effective sample size (ESS), an approximation to the number of ``independent" draws in MCMC sampling. It shows that the ESS of all parameters is greater than $400$, which is considered to be adequate by \cite{vehtari2021rank}. The ESS of $\tau$ is significantly lower than that of $\beta$, a possible reason is that the frailty random effect might be time dependent $w_i(t)$ rather than a time-fixed effect. 
Another metric of interest is the average time needed to generate each effective sample, called MCMC Pace. Stan develop team emphasized the importance of MCMC Pace, and the definition is given by the team of NIMBLE in \cite{nimble} as the time consuming of generating one effective sample. The MCMC Pace to generate $\tau$ is much higher than that of $\beta$, and we conjecture the possible reason is that the posterior distribution has a long upper tail leading to outliers in posterior samples, which slows down the speed to generate effective samples.

\subsection{Parametric estimation \RNum{2}: simulation}
Another simulation study is considered to evaluate the performance of parametric estimation of the MCMC procedure. Our simulation aims to simulate the occurrences of multiple events on the same individual. We take $K=2$ and $G=3$ denoting the number of types of events and number of treatment strata, respectively. The simulation includes two independent covariates,  $x_i \sim \text{Bin}(1, 0.5)$ and $x_2 \sim N(0, 1)$ to incorporate indicator variable and continuous variable as well. For $k =1, 2, j=1, 2, 3$, the baseline survival functions $S_{0kj}$ are set as:
\begin{itemize}
    \item $S_{011} = 1-0.5(LN(-0.25, 1)+LN(0.25, 1))$;
    \item $S_{012} = 1-0.5(LN(-0.5, 1)+LN(0.65, 1))$;
    \item $S_{013} = 1-0.5(LN(-0.65, 1)+LN(1.25, 1))$;
    \item $S_{021} = 1 - LN(0, 1)$; 
    $S_{022} = 1 - LN(-0.5, 1)$;
    $S_{023} = 1 - LN(0.5, 1)$
\end{itemize}
When $k=1$, the three baseline survival functions are crossed whereas when $k=2$, the three curves are not. 
The vector of regression coefficients is $\beta = (1, 1)^T$ and the log frailty random effect $v_i \sim N(0, 1)$ independently. 
The survival time is generated following model \eqref{basicmod}. The censoring variable of each event is generated from $\text{Unif}(4, 6)$ independently, leading to a censoring rate of about $28\%$. We set the number of subjects to be $90$ and they are equally divided into three treatment strata. We repeat the simulation for $150$ times. 

Table \ref{tabsim} summarizes the results for regression parameters $\beta$ and the standard deviation of frailty effect $\tau$, including the averaged bias (BIAS), root of mean square error (RMSE), posterior estimated standard deviation (ESD) of each point estimate (posterior mean for $\beta$ and median for $\tau$), the standard deviation (across 150  replicated simulations) of the point estimate (SDE), and the coverage probability (CP) of the 95\% credible interval (given by Wald-type credible interval). The results show that the point estimates of $\beta$ and $\tau$ have quite little bias with low RMSE, ESD values are close to the corresponding SDEs, and the CP values are close to the nominal 95\%. 

\begin{table}[ht]
\centering
\caption{Simulation results for the parametric terms. BIAS, averaged bias among the 150 simulations; RMSE, root of mean square error of the estimation; ESD, averaged posterior estimated standard deviation; SDE, the standard deviation of point estimate; CP, the coverage probability of 95\% credible interval. }
\begin{tabular}{ccccccc}
  \hline
Parameter& BIAS & RMSE & ESD & SDE & CP \\ 
  \hline
      $\beta_1=1$ & -0.062 & 0.042 & 0.222 & 0.196 & 96.7\\ 
 $\beta_2=1$ &-0.025 & 0.023 & 0.148 & 0.152 & 92.7 \\ 
      $\tau=1$ & -0.078 & 0.056 & 0.213 & 0.224 & 96.7 \\ 
   \hline
\end{tabular}
\label{tabsim}
\end{table}

\begin{backmatter}
\section{Discussion}
In this chapter, we show the power of Bayesian computing illustrated by successfully applying the ANOVA DDP model as the nonparametric prior for a relatively complicated shared frailty model.
Our survival-function based version of the ANOVA DDP,  modified based on the ANOVA DDP directly in subsection 3.3, is constructed for the shared frailty model, but can reduce to modeling the univariate dependent survival functions by involving the continuous covariates into the predictor space of the ANOVA DDP. Hence, our work is an extension of \cite{de2009bayesian} to some extent. 
However, the proposed GSFM is different to the Linear DDP models for generalization of accelerated failure time model (\cite{hanson2013surviving, riva2021survival}). 
Furthermore, although we point out that there exists potentially dual dependence for dual stratification of treatment strata and recurrences, we just simply allow dependence in treatment strata and assume that the recurrences are independent in our methodology demonstration. 
The dependence across recurrences per subject is dealt with only by the parametric frailty random effect in the proposed shared frailty model. It is more reasonable to be incorporated into the baseline survival functions so that the interaction effects between recurrence and treatment may be accounted for.
Under the one-level stratification, \cite{hanson2012bayesian} modeled such serial correlation among baseline hazard functions by constructing the so-called dependent tail free process as the prior. It is non-trivial to accommodate dual temporal and stratified dependency as a future research plan.


\section*{Acknowledgments}
Chong Zhong’s research was partially supported by GRF1531519, RGC, HKSAR.
Zhihua Ma's research was partially supported by Shenzhen Institutions Stability Support Program 20200812101943002, China.
Junshan Shen's research is partially supported by the Beijing Natural Science Foundation 1192006, China. 
Catherine Liu's research was partially supported by HKPOLYU grant YBTR, and GRF1531519, RGC, HKSAR.

\section*{Thanks}
The first author Chong Zhong owes deep thanks to his parents. 
The authors thank Miss. Lulu Zhang for her efficient technical supports in tex and figures. 
The authors thank the service manager Ms. Romina Rovan for her courtesy and professional service. 
The authors thank the invitation from the editor. 
\begin{authordetails}
	
	
	\author{Chong Zhong$^{1\dagger}$, Zhihua Ma$^{2\dagger}$, Junshan Shen$^{3\dagger}$, Catherine Liu$^{1*\dagger}$}
	\address[1]{The Hong Kong Polytechnic University, Hong Kong, China}
	\address[2]{Shenzhen University, Guangdong, China}
	\address[3]{Capital University of Economics and Business, Beijing, China}
	%
	
	\IntechOpentext{\textcopyright\ \the\year{} The Author(s). License IntechOpen. This chapter is distributed under the terms of the Creative Commons Attribution License (http://creativecommons. org/licenses/by/3.0), which permits unrestricted use, distribution, and reproduction in any medium, provided the original work is properly cited.}
	
	
\end{authordetails}
\end{backmatter}
\bibliography{main}

\end{document}